\documentclass[showpacs,preprintnumbers,prd,nofootinbib,floats,amssymb,floatfix]{revtex4}
\usepackage{amsmath}
\usepackage{amsfonts}
\usepackage{graphicx}
\usepackage{hyperref}
\usepackage{amsmath}
\usepackage{amstext}
 
\begin{document}

\title {Dirac's  and Generalized Faddeev-Jackiw brackets for  Einstein's theory in   the $G \rightarrow 0$ limit}
 \author{ Alberto Escalante}  \email{aescalan@ifuap.buap.mx}
 \affiliation{Instituto de F{\'i}sica Luis Rivera Terrazas, Benem\'erita Universidad Aut\'onoma de Puebla, (IFUAP). Apartado
 postal J-48 72570 Puebla. Pue., M\'exico}
\affiliation{LUTh, Observatoire de Paris, Meudon, France}
 \author{Omar Rodr{\'i}guez Tzompantzi} 
 \affiliation{ Facultad de Ciencias F\'{\i}sico-Matem\'{a}ticas, Benem\'erita Universidad Au\-t\'o\-no\-ma de Puebla,
 Apartado postal 1152, 72001 Puebla, Pue., M\'exico}
 
\begin{abstract}
In this paper  the Dirac  and Faddeev-Jackiw formulation  for  Einstein's theory in   the $G \rightarrow 0$ limit is performed;   the fundamental   Dirac's  and Faddeev-Jackiw brackets for the theory are obtained. First,   the Dirac brackets are  constructed  by  eliminating the second class constraints remaining the first class ones, then  we fix the gauge and we convert the first class constraints into  second class constraints and the new  fundamental Dirac's brackets are computed. Alternatively, we reproduce all  relevant Dirac's results by means of the symplectic method. We identify the Faddeev-Jackiw constraints  and we prove that the Dirac and the Faddeev-Jackiw brackets coincide to each other. 
\end{abstract}

\date{\today}
\preprint{}
\maketitle
\section{INTRODUCTION}
The study of gauge theories by means of the canonical analysis  developed by Dirac  is an important step that should be performed. In fact, it is well-known that the correct identification of the constraints and the symplectic structure of a singular  system  is the best guideline for performing its  quantization \cite{1a, 2a}. In this respect, Dirac's formalism  is an elegant framework  for studying  singular systems, which allow us to know   the constraints, the fundamental Poisson brackets, the extended hamiltonian and  the Dirac brackets  \cite{3, 4a}. However, in some cases,  to develop   the Dirac method is   large and tedious task  and some times if   all the Dirac steps are not  applied correctly or  some of them are   omitted  \cite{5, 6}, then    the obtained results could be incorrect  \cite{7, 8, 20x}. Hence, because of  these complications,   it is necessary to use alternative formulations that could give us a complete description of the theory, in this sense,    there is a different  approach  for studying  gauge theories, the  so-called  the Faddeev-Jackiw  [FJ] formalism \cite{9}. The [FJ] method   is a symplectic approach, namely, all the relevant information of the theory can be obtained through an invertible symplectic tensor, it is  constructed by means  the symplectic variables that are identified as the degrees of freedom.  Since  the theory under interest is singular  there will be constraints,  and [FJ] has the   advantage that  all the  constraints of the theory are treated  at the same footing, namely,   it is not necessary  to perform  the  classification of the constraints in primary, secondary, first class or second class such  as in Dirac's method is done \cite{10}. Furthermore,  in [FJ] approach   it is possible to obtain the gauge transformations of the theory and the generalized [FJ]  brackets  coincide  with the Dirac  ones, basically in [FJ]  we  only choose the symplectic variables either the configuration space or the phase space and by fixing the appropriated gauge,   we can invert the symplectic matrix in order to obtain a complete analysis.  In this manner, it is possible to obtain all the Dirac results,  say, in Dirac's approach  we can  construct the Dirac brackets  by means two ways;  eliminating  only the second class constraints  remaining the first class ones or we can fix the gauge and  convert  the first class constraints into second class ones, in any case, we can reproduce these Dirac's results by means  of  the [FJ] framework. In fact, for the former  we use the configuration space as symplectic variables, for the later we use the phase space \cite{11, 12}.\\
With these motivations  we study the Einstein theory in   the $G \rightarrow 0$ limit.  Einstein's theory in   the $G \rightarrow 0$ limit is an interesting theory, it   has as  scenario to Minkowski spacetime,   lacks of physical degrees of freedom   with  reducible constraints; in certain sense, it is a copy of  a four dimensional $BF$ theory. The theory has been analyzed in the context of Dirac by  introducing a kind of ADM variables  \cite{13},  the analysis shows that there are only reducible first class constraints and the algebra of the constraints is similar  to the algebra between the first class constraints found in  General Relativity [GR].  On the other side, the theory  has been also analyzed in \cite{14} by using a pure  Dirac's  approach,  namely, the analysis was developed  by following all the Dirac steps and using   the full phase space as degrees of freedom without introducing extra variables. In  that  paper was reported that the theory has reducible  first class constraints and irreducible second class constraints, the algebra of the constraints is closed and it has the required structure that Dirac's framework demands \cite{8}, namely, the Poisson brackets between first class constraints are  linear in first class constraints and quadratic  in second class constraints etc., \cite{2a}. Nevertheless,  in spite of those analysis  the Dirac brackets were  not reported. In fact, if we use the results given in \cite{13}, it is difficult develop  the construction of the Dirac brackets in terms of   ADM type variables. On the other hand,  in the paper  reported in \cite{14}, there are first class and second class constraints,  thus, in order to construct the Dirac brackets we can choose fixing  or not fixing  the gauge, but the reducibility among the constraints complicate the computation  and it is necessary to expand the phase space with canonical auxiliary fields   for obtaining that aim. Thus, in this paper we use  the results reported in \cite{14} and we construct the Dirac brackets by fixing or not fixing the gauge, then we perform the [FJ] analysis and we obtain by  a different way all  relevant Dirac's results.\\
The paper is organized as follows: In section I, we develop a review of the results obtained in \cite{14}, then we construct the Dirac brackets for the theory under study by eliminating only the second class constraints. In section II, we reproduce the Dirac results obtained in the previous section by using the [FJ] approach. We will work with the configuration space as symplectic variables, in  order to invert the symplectic tensor we fixing the temporal gauge, then we prove that the generalized [FJ]  and the Dirac  brackets are the same. In section III, we use the fact that  the scenario of the theory under study corresponds to Minkowski spacetime background,  the Dirac brackets are constructed  by fixing the gauge;  we convert to the first class constraints into second class constraints. In order to invert the matrix whose entries are given by the Poisson brackets between the second class constraints,   we expand the phase space by introducing auxiliary canonical fields  that will be  useful for constructing the Dirac brackets, then the fundamental Dirac's brackets are calculated. In section IV, we reproduce the results obtained in the Section III by using the [FJ] analysis. In fact, now   we  use  the phase space as symplectic variables,   we fix the gauge by using  a Coulomb gauge, because of the reducibility among the first class constraints  we also  introduce a reducible gauge.  Then we show that the generalized [FJ] and Dirac's brackets coincide to each other. Finally,  in Section V we present some remarks and conclusions.
\section{Hamiltonian analysis}
The action that  we will study in this section is given by Einstein's theory of gravity written in the first order formalism expressed by \cite{13, 14, 15}
\begin{equation}
S[A, e]= \frac{1}{8}  \int  \epsilon^{\alpha \beta \gamma \delta} \epsilon^{IJKL} (e_{\alpha I} e_{\beta J}R_{\gamma \delta KL})  dx^4,
\label{1}
\end{equation}
where $\epsilon^{IJKL}$  is the completely antisymmetric object with $\epsilon^{0123}=1$, $e_{\alpha}^I$ is the tetrad field and $R_{\gamma \delta KL}= \partial_\gamma A_{\delta KL} - \partial_\delta  A_{\gamma  KL} + G\left(A_{\gamma K}{^{J}} A_{\delta JL} -A_{\delta K}{^{J}} A_{\gamma JL} \right) $  is the curvature of the $SO(3,1)$ connection $A_\alpha{^{IJ}} $. Here, $G$ is the  gravitational coupling constant, $\mu,  \nu=0, 1, ...,3$ are space-time indices, $x^\mu$ are the coordinates that label the points for the 4-dimensional manifold $M$ and $I, J=0,1,...,3$ are internal indices that can be raised and lowered by the internal Lorentzian metric $\eta^{IJ}= (1, -1, -1, -1)$.
In the reference \cite{13} is reported that by setting the  $G\rightarrow 0$ limit and performing a change of variables,  the action (\ref{1}) is reduced to a copy of   $BF$-like theory 
\begin{equation}
S[e, B]= \frac{1}{2} \int \epsilon^{\alpha \beta \gamma \delta} B_{\alpha \beta}^I \left(\partial_{\gamma}e_{\delta I}- \partial_{\delta}e_{\gamma I} \right) dx^4,
\label{eq2}
\end{equation}
where $B_{\alpha \beta}^I= -\frac{1}{2}\epsilon^{IJKL} e_{[\alpha J} A_{\beta] KL} $   and the fields $e_\alpha^I$ are a collection of four   gauge invariant $U(1)$ vector fields. It is important to comment that one of the equations of motion obtained from (\ref{eq2})  given by $\epsilon^{\alpha \beta \gamma \delta} \partial_ \gamma e_\delta ^I=0$,  implies that $e_{\alpha}^I = \partial_\alpha f^I$, so  the metric $g_{\mu \nu}= \partial_\mu f^I  \partial_\nu f^J \eta_{IJ} $ corresponds locally to Minkowski spacetime, this fact will be used in the following sections. Furthermore,  in \cite{13} it was performed the Hamiltonian analysis  of the action  (\ref{eq2}) by using a  kind of ADM variables and working on a reduced  phase space. In that paper  only first class constraints  were identified  and   the   construction of the Dirac brackets of theory was  not reported.   On the other hand,  in \cite{14}  it  was reported a pure Hamiltonian analysis, using    the full phase  space, and   were identified reducible first class constraints and irreducible  second class constraints. Furthermore,   the extended action  and  the complete  Poisson algebra among  the first class constraints were reported.  In this manner, the structure of the constraints presented in \cite{14} is more suitable to work  than the structure of the constraints reported in \cite{13}, in fact, by using the results reported in \cite{14} there are two possibilites for constructing the Dirac brackets, by fixing or not fixing the gauge; in this work,  we will use the results reported in \cite{14} for constructing the Dirac brackets and then we perform the  [FJ] analysis,  we shall  prove that  the Dirac brackets and the generalized [FJ] brackets are the same. \\
Therefore, by working a pure Dirac's analysis  of the action (\ref{eq2}) \cite{14},    the following extended action was reported 
\begin{eqnarray}
S[ e_\mu^I, \Pi ^{\mu}_I, B_{\mu \nu}^I, \Pi^{\mu \nu} _I, u_0^I, u^I, u_{0a}^I, u_a ^I, v_a ^I, v_{ab} ^I  ]&=& \int \Big \{ \dot{e}_\mu^I \Pi ^{\mu}_I - H_E -  v_a ^I  \chi_I ^a - v_{ab}^I  \chi^{ab}_I  \Big\}, 
\end{eqnarray}
where $v_a ^I, v_{ab}^I  $  are Lagrange multipliers enforcing the second class constraints $ \chi_I ^a$ and $ \chi^{ab}_I $ given by 
\begin{eqnarray}
\chi^{a}{_{I}}&=&\Pi^{a}{_{I}}-\eta^{abc}B_{Ibc} \approx 0, \nonumber \\
\chi^{ab}{_{I}}&=&\Pi^{ab}{_{I}} \approx 0, 
\label{eq4}
\end{eqnarray}
and $H_E$ is the extended Hamiltonian given by 
\begin{equation}
H_E= - B_{0a}^I \left[  \eta^{abc} (\partial_{b} e_{Ic}-\partial_{c} e_{Ib})-2\partial_{b}\Pi^{ab}{_{I}} \right] - e_0^I \partial_a \Pi^a_I - u_0^I\gamma^0_I - u^I \gamma_I - u_a^I \gamma^a_I - u_{0a}^I \gamma^{0a}_I,
\end{equation}
with $u_0^I, u^I, u_{0a}^I, u_a ^I$ being Lagrange multipliers enforcing the following first class constraints 
 \begin{eqnarray}
\gamma^{0}{_{I}}&=&\Pi^{0}{_{I}} \approx 0, \nonumber \\
\gamma^{0a}{_{I}} &=&\Pi^{0a}{_{I}} \approx 0, \nonumber \\
\gamma_I &=& \partial_a \Pi^a_I ,\nonumber \\
\gamma^{a}{_{I}}&=&\eta^{abc}(\partial_{b} e_{Ic}-\partial_{c} e_{Ib})-2\partial_{b}\Pi^{ab}{_{I}} \approx 0.
\label{eq3}
\end{eqnarray}
We can observe that  the price to pay for working with the complete phase space, is that there are second class constraints, which makes a difference with respect  to the results  reported in \cite{13} where only   first class constraints were  found. Moreover,  we also observe that these constraints are not independent because there exist reducibility among the constraints given by $\partial_a \gamma^{a}{_{I}}=0$, which   complicates   the construction of the Dirac brackets, however, that trouble can be fixed by enlarging the phase space as we will see below.  Thus, in order to  construct the Dirac brackets without fixing the gauge,  we calculate  the Poisson brackets among the second class constraints  
\begin{eqnarray}
\{\chi^{a}{_{I}}(x),\chi^{b}{_{J}}(y)\}&=&0, \nonumber \\
\{\chi^{a}{_{I}}(x),\chi^{bc}{_{J}}(y)\}&=& -\eta^{aij}\frac{\delta^{bc}{_{ij}}}{2}\eta_{IJ}, \nonumber \\
\{\chi^{ef}{_{I}},\chi^{ab}{_{J}}(y)\}&=&0.\nonumber 
\label{eq5}
\end{eqnarray}
Furthermore,  the matrix, namely $C_{\alpha \beta}$,  whose entries are given by the Poisson brackets among the second class constraints  takes the form
\begin{eqnarray*}
\label{eq}
C_{\alpha \beta}=
\left(
  \begin{array}{ccccc}
   0   &   -\eta^{abc}\eta_{IJ}                                                                           \\
    \eta^{abc}\eta_{IJ}  &     0                                                                        \\
      \end{array}
\right), 
\end{eqnarray*}
and its inverse is given by 
\begin{eqnarray*}
\label{eq}
(C^{\alpha \beta})^{-1}=
\frac{1}{2}\left(
  \begin{array}{ccccc}
   0   &   \eta_{abc}\eta^{IJ}                                                                          \\
    \ -\eta_{abc}\eta^{IJ}  &     0                                                                        \\
      \end{array}
\right). 
\end{eqnarray*}
In this manner, the Dirac bracket of two functionals $F$, $G$ defined on the phase space,  is expressed by
\[
\{F(x),G(z)\}_{D}\equiv\{F(x),G(z)\} - \int d^{2}ud^{2}w\{F(x),\xi_{\alpha}(u)\}(C{^{\alpha\beta}})^{-1}\{\xi_{\beta}(w),G(z)\},
\]
where $\{F(x),G(z)\}$ is the Poisson bracket  between two functionals $F,G$,  and $\xi_{\alpha}= ( \chi_I ^a,  \chi^{ab}_I)$ represent  the set of  second class constraints. By using this fact, we obtain the following  Dirac's brackets 
\begin{eqnarray}
\{e^{I}{_{a}}(x), \Pi^{b}{_{J}}(y)\}_{D}&=&\delta^{b}{_{a}}\delta^{I}{_{J}}\delta^{3}(x-y), \nonumber \\
\{\Pi^{a}{_{I}}(x),\Pi^{b}{_J}(y)\}_{D}&=&0,\nonumber\\
\{e^{I}{_{a}}(x),e^{J}{_{b}}(y)\}_{D}&=&0,\nonumber \\
\{e^{I}{_{a}}(x),\Pi^{gd}{_{J}}(y)\}_{D}&=&0, \nonumber\\
\{\Pi^{a}{_{I}}(x),B^{J}{_{cd}}(y)\}_{D}&=&0, \nonumber \\
\{\Pi^{a}{_{I}}(x),\Pi^{cd}{_{J}}(y)\}_{D}&=&0, \nonumber \\
\{e^{I}{_{a}}(x),B^{J}{_{gd}}(y)\}_{D}&=&\frac{\eta_{agd}}{2}\eta^{IJ}\delta^{3}(x-y), \nonumber\\
\{B^{I}{_{ab}}(x),\Pi^{gd}{_{J}}(y)\}_{D}&=&0.
\label{eq5}
\end{eqnarray}
Now, by using these  Dirac's brackets we can  show that the algebra among the constraints vanish identically. Hence, there is a difference respect the results  obtained in \cite{13} because  in that work  it was used a kind of  ADM variables and the Poisson algebra among the constraints is linear combinations of constraints. In this section, we have worked with the full phase space without resorting to extra variables and the algebra among the constraints  is more suitable for working with them. Furthermore, we have eliminated only the second class constraints, and  at this step it  was not necessary to take into account the reducibility conditions; however, if we fix the gauge and we convert  the fist class constraints into second class constraints, then the  reducibility conditions will be taken into account, we shall  explain this fact in later sections. \\

\section{Faddeev-Jackiw Formalism by using a temporal gauge }
Now we will reproduce the above results by using the [FJ] formalism. We can see that  the action (\ref{eq2}) can be written in the following form
\begin{eqnarray}
{\mathcal{L}}{^{(0)}}&=&\eta^{abc}B_{Ibc}\dot e^{I}{_{a}}-V^{(0)},
\label{eq8}
\end{eqnarray}
where the symplectic potential is given by  $V^{(0)}=-\eta^{abc}B_{I0a}(\partial_{b}e^{I}{_{c}} - \partial_{c}e^{I}{_{b}}) - \partial_{c}(\eta^{abc}B_{Iab})e^{I}{_{0}}$.  The corresponding symplectic equations of motion are given by \cite{9}
\begin{equation}
f^{(0)}_{ij}\dot{\xi}^{j}=\frac{\partial V^{(0)}(\xi)}{\partial\xi^{i}},
\label{eq4}
\end{equation}
where the symplectic matrix $f^{(0)}_{ij}$ takes the form
\begin{equation}
f^{(0)}_{ij}(x,y)=\frac{\delta a_{j}(y)}{\delta\xi^{i}(x)}-\frac{\delta a_{i}(x)}{\delta\xi^{j}(y)},
\label{4}
\end{equation}
with $\xi{^{(0)i}}$ and $a{^{(0)}}{_{i}}$ representing  a set of symplectic variables. Hence, in order to reproduce by means of the  [FJ] method the  results obtained in previous section  we will work by using the configuration space as symplectic variables. In fact,  it has been  showed in \cite{12}  that if we construct the Dirac brackets by eliminating only the second class constraints, then in the [FJ] scheme it is necessary to work with the configuration space.  We know that introducing the Dirac brackets by  eliminating  the second class constraints, in particular  eliminating  the primary second class constraints, then the momenta become to be  a label because the momenta can be expressed in terms of the fields.   Therefore,  from (\ref{eq8}) we identify the following symplectic variables\quad $\overset{(0)}{\varepsilon}{^{i}}=(e^{I}{_{a}},B_{I0a},e^{I}{_{0}},B^{I}{_{ab}})$ and the 1-forms \, $a^{(0)}{_{i}}=(\eta^{abc}B_{Ibc},0,0,0)$. In this manner,  by using these symplectic variables, we construct the following symplectic matrix

\begin{eqnarray}
\label{eq}
\overset{(0)}{f}_{ij}=
\left(
  \begin{array}{cccc}
  0   &\quad   0  &\quad   0     &\quad   -\eta^{abc}\eta_{IJ}                                                                        \\
  0 &\quad     0   &\quad  0   &\quad   0                                                                        \\
  0   &\quad  0   &\quad   0      &\quad   0                                                                     \\
  \eta^{abc}\eta_{IJ}   &\quad   0   &\quad   0    &\quad  0                                                                  \\ 
 \end{array}
\right) \delta^{3}(x-y).
\label{eq11s}
\end{eqnarray}\\
We observe that the matrix (\ref{eq11s}) is not invertible, hence, there exist constraints. It is easy to observe that there are the following modes  \, $\overset{(0)}v_{1}=(0,v^{B^{I}_{0a}},0,0)$ and $\overset{(0)}v_{2}=(0,0,u^{e^{I}{_{0}}},0)$ where $v^{B^{I}_{0a}}$ and $u^{e^{I}{_{0}}}$ are arbitrary functions. By using these modes we obtain the following [FJ] constraints
\begin{eqnarray}
\overset{(1)}{\Omega}{^{aI}}&=&\int d^{2}x(v^{(0)})^{T}_{i}(x)\frac{\delta}{\delta\xi^{(0)i}(x)}\int d^{2}y V^{(0)}(\xi) \nonumber \\
            &=&\eta^{abc}(\partial_{b}e^{I}{_{c}} - \partial_{c}e^{I}{_{b}})=0,
    \label{eq10}
\end{eqnarray}
\begin{eqnarray}
\overset{(2)}{\Omega}{_{I}}&=&\int d^{2}x(v^{(0)})^{T}_{i}(x)\frac{\delta}{\delta\xi^{(0)i}(x)}\int d^{2}y V^{(0)}(\xi) \nonumber \\
            &=& \partial_{c}(\eta^{abc}B_{Iab})=0.
          \label{eq11}
\end{eqnarray}
We can observe  that (\ref{eq10}) is a reducible constraint because of $\partial_a \overset{(1)}{\Omega}{^{aI}}=0$.  Now,  we will  observe  if there are present more constraints in the [FJ]  context. For this aim, we write in matrix form the following system \cite{9, 10}
\begin{equation}
   f^{}_{kj}\dot{\xi}^{j}=Z_{k}(\xi),
\label{eq14a}
\end{equation}
where
\begin{eqnarray}
Z_{k}(\xi)=
\left(
 \begin{array}{cccc}
   \frac{\partial V^{(0)}(\xi)}{\partial \xi^{i}}\\
   0\\
   0\\
  \end{array}
\right),
\label{eq15a}
\end{eqnarray}
and
\begin{eqnarray}
f^{}_{kj}=
\left(
 \begin{array}{cccc}
   f^{(0)}_{ij}\\
   \frac{\partial\Omega^{(1)}}{\partial\xi^{i}}\\
   \frac{\partial\Omega^{(2)}}{\partial\xi^{i}}\\
  \end{array}
\right)=\left(
 \begin{array}{cccccc}
   0&0&0&-\eta^{abc} 	\eta_{IJ}&\\
   0&0&0&0&\\
   0&0&0&0&\\
   \eta^{abc} 	\eta_{IJ}&0&0&0&\\
  2 \eta^{abc}\delta_{I}^{J}\partial_{b}&0&0&0&\\
   0&0&0&\eta^{abc}\delta_{I}^{J}\partial_{b}&\\
  \end{array}
\right)\delta^3(x-y),
\label{eq16aa}
\end{eqnarray}
we  can observe that (\ref{eq16aa})  is  not a square matrix as expected, however, it has   linearly independent modes. It is straightforward calculate the modes, say $(v^{(1)})_{k}^{T}$,  and  from the  contraction $(v^{(1)})_{k}^{T}Z_{k}=0$ we can prove that    there are not more [FJ] constraints. Now, we introduce all that  information  by constructing a new symplectic Lagrangian. For this aim, we use the Lagrange multipliers $\lambda_{aI}$ and $\rho^{I}$ associated to the constraints,  the new symplectic Lagrangian is given by
\begin{eqnarray}
{\mathcal{L}}{^{(1)}}&=&\eta^{abc}B_{Ibc}\dot e^{I}{_{a}} - \overset{(1)}{\Omega}{^{aI}}\dot \lambda_{aI} - \overset{(2)}{\Omega}{_{I}}\dot \rho^{I} - V^{(1)},
\label{eq12}
\end{eqnarray}
where \, $V^{(1)}=V^{(0)} \mid _{\overset{(1)}{\Omega},\overset{(2)}{\Omega}=0} =0$. In this manner,  the symplectic Lagrangian is given by 
\begin{eqnarray}
{\mathcal{L}}{^{(1)}}&=& \eta^{abc}B_{Ibc}\dot e^{I}{_{a}} - \overset{(1)}{\Omega}{^{aI}}\dot \lambda_{aI} - \overset{(2)}{\Omega}{_{I}}\dot \rho^{I}, 
\label{eq13a}
\end{eqnarray}
from (\ref{eq13a}) we identify the following symplectic variables given by\\ ${\overset{(1)} \varepsilon}{^{i}}=(e^{I}{_{a}},\lambda_{aI},B_{Ibc},\rho^{_{I}})$ and the 1-forms\, $\overset{(1)}a{_{i}}=(\eta^{abc}B_{Ibc},-\overset{(1)}{\Omega}{^{aI}},0,-\overset{(2)}{\Omega}{_{I}})$. By using these symplectic variables we obtain the following symplectic matrix
\begin{eqnarray}
\label{eq}
\overset{(1)}{f}{_{ij}}=
\left(
  \begin{array}{cccc}
  0   &\quad   -2\delta^{I}{_{J}}\eta^{abc}\partial_{b}  &\quad   -\eta^{abc}\delta^{I}{_{J}}  &\quad 0                                                                        \\
  2\delta^{I}{_{J}}\eta^{abc}\partial_{b}  &\quad     0   &\quad  0   &\quad   0   
\\
  \eta^{abc}\delta^{I}{_{J}}   &\quad  0   &\quad   0      &\quad  -\delta^{I}{_{J}}\eta^{abc}\partial_{c}                                                                      \\
  0 &\quad  0   &\quad   \eta^{abc}\delta^{I}{_{J}}\partial_{c}    &\quad  0             \\ 
 \end{array}
\right) \delta^{3}(x-y),
\label{eq20}
\end{eqnarray}
were we can see that $\overset{(1)}{f}{_{ij}}$ \,is a singular matrix. In fact, this  matrix has 40 null vectors, thus,   it is not invertible, however, we have showed that there are not more constraints. Therefore, in the scheme  this means that the theory has a gauge symetry.
In order to invert the symplectic matrix (\ref{eq20}),  we fix the following temporal  gauge\, $e^{I}{_{0}}=0$ and $B_{I0a}=0$, this fact  means that $\rho^{I}=$constant and $\lambda{_{aI}}=$constant. In this manner,  we introduce this information  in a new symplectic Lagrangian given by 
\begin{eqnarray}
{\mathcal{L}}{^{(2)}}&=& \eta^{abc}B_{Ibc}\dot e^{I}{_{a}} - (\overset{(1)}{\Omega}{^{aI}} - \alpha{^{aI}})\dot \lambda_{aI} - (\overset{(2)}{\Omega}{_{I}}-\rho_{I})\dot \phi^{I}, 
\label{eq13}
\end{eqnarray}
where we identify the following symplectic variables ${\varepsilon}{^{i}}=(e^{I}{_{a}},\lambda_{aI},\phi^{I},B_{Ibc},\alpha^{aI},\rho_{I})$ \, and the 1-forms\, $a_{i}= (\eta^{abc}B_{Ibc},-(\overset{(1)}{\Omega}{^{aI}} - \alpha{^{aI}}),-(\overset{(2)}{\Omega}{_{I}}-\rho{_{I}}),0,0,0)$. By using these symplectic variables, the symplectic matrix is given by
\begin{eqnarray}
\label{eq16a}
\overset{(2)}{f}_{ij}=
\left(
  \begin{array}{cccccc}
  0   &\quad   -2\delta^{I}{_{J}}\eta^{abc}\partial_{b}  &\quad   0     &\quad   -\eta^{abc}\delta^{I}{_{J}}  &\quad 0   &\quad 0                                                                        \\
  2\delta^{I}{_{J}}\eta^{abc}\partial_{b}  &\quad     0   &\quad  0   &\quad   0   &\quad  -\delta^{a}{_{b}}\delta^{I}{_{J}}  &\quad  0                                                                     \\
  0   &\quad  0   &\quad   0      &\quad   -\delta^{I}{_{J}}\eta^{abc}\partial_{c}  &\quad   0      &\quad   \delta^{I}{_{J}}                                                              \\
  \eta^{abc}\delta^{I}{_{J}}   &\quad   0   &\quad   \delta^{I}{_{J}}\eta^{abc}\partial_{c}     &\quad  0     &\quad  0  &\quad  0                                                                  \\ 
0  &\quad   \delta^{a}{_{b}}\delta^{I}{_{J}}  &\quad  0   &\quad  0   &\quad  0   &\quad  0 \\
0 &\quad   0  &\quad   -\delta^{I}{_{J}}   &\quad  0 &\quad 0    &\quad  0\\
 \end{array}
\right) \delta^{3}(x-y). \nonumber \\
\end{eqnarray}
We can observe that $f^{(2)}_{ij}$ is not singular, therefore we can construct its inverse. The inverse of (\ref{eq16a}) is called    the symplectic tensor 
\begin{eqnarray}
\label{eq17a}
(\overset{(2)}{f}_{ij}){^{-1}}=
\left(
  \begin{array}{cccccc}
  0   &\quad   0  &\quad   0     &\quad  \frac{\eta_{abc}}{2}\delta^{I}{_{J}}  &\quad 0   &\quad   \delta^{I}{_{J}}                                                                       \partial_{a} \\
  0  &\quad     0   &\quad  0   &\quad   0   &\quad  \delta^{I}{_{J}}\delta^{a}{_{b}}  &\quad  0                                                                     \\
  0   &\quad  0   &\quad   0      &\quad   0  &\quad   0      &\quad   -\delta^{I}{_{J}}                                                              \\
 -\frac{ \eta_{abc}}{2}\delta^{I}{_{J}}   &\quad   0   &\quad   0  &\quad  0     &\quad  -\delta^{I}{_{J}}\delta^{ad}_{bc}\partial_{d}  &\quad  0                                                                  \\ 
0  &\quad   -\delta^{I}{_{J}}\delta^{a}{_{b}}  &\quad  0   &\quad  \delta^{I}{_{J}}\delta^{ad}_{bc}\partial_{d}   &\quad  0   &\quad  0 \\
-\delta^{I}{_{J}}\partial_{a}  &\quad   0  &\quad   \delta^{I}{_{J}}   &\quad  0 &\quad 0    &\quad  0\\
 \end{array}
\right) \delta^{3}(x-y).
\end{eqnarray}\\
In this manner, we identify the following [FJ] generalized brackets
\begin{align*}
\{e^{I}{_{a}},B_{Jbc}\}_{FJ}&=\delta^{I}{_{J}}\frac{\eta_{abc}}{2}\delta^{3}(x-y),\\
\{e^{I}{_{a}},\rho{_{J}}\}_{FJ}&=\delta^{J}{_{I}}\partial_{a}\delta^{3}(x-y),\\
\{\lambda_{aI},\alpha^{bJ}\}_{FJ}&=\delta^{b}{_{a}}\delta^{J}{_{I}}\delta^{3}(x-y),\\
\{\phi^{I},\rho_{J}\}_{FJ}&=-\delta^{I}{_{J}}\delta^{3}(x-y),\\
\{B_{Iab},\alpha^{cJ}\}_{FJ}&=-\delta^{J}_{I}\delta^{cd}{_{ab}}\partial_{d}\delta^{3}(x-y), 
\end{align*}
where we can see that these [FJ] brackets  coincide with  the  Dirac  brackets found in previous section. Furthermore, we have not taken into account the reducibility conditions just like it  was done in Dirac's method, but  in latter sections we will. 
\section{Dirac's brackets by fixing the gauge}
It is well-known  that in [GR] the coordinates of the space and time lacks of physical meaning, the physical relevance in [GR] is giving by  the relation of fields respect to other fields \cite{16}. In this respect, we can not localise the gravitational field  in the spacetime (gauge fixing) because the spacetime is a dynamical system, however, for the theory under study  we have commented above  that one of the equations of motion obtained from  the action (\ref{eq2}) implies that the spacetime is locally Minkowski. Therefore, we can fix the gauge and we will construct the Dirac brackets in order to perform the quantization of the theory. Hence, for this aim, we  fix the following gauge $e^{I}{_{0}} \approx 0$,  $B^{I}{_{0a}} \approx 0$, $\partial^{a}e^{I}{_{a}} \approx 0$ and $-\partial^{b}B_{ab}{^{I}}  \approx 0$. We can observe that the gauge $-\partial^{b}B_{ab}{^{I}}$ is also reducible, and this fact does not allow us calculate the Dirac's brackets, however, we will introduce auxiliary fields in order to convert the reducible constraints in irreducible ones  \cite{17, 18}.   Thus  we obtain the following set of second class constraints 
\begin{align} 
\chi_{1}&=e^{I}{_{0}} \approx 0, &
\chi_{6}&=\partial_{a}\Pi^{a}{_{I}} \approx 0, \nonumber \\
\chi_{2}&=\Pi^{0}{_{I}} \approx 0, &
\chi_{7}&=-\partial^{b}B_{ab}{^{I}} + \partial_{a}q^{J} \approx 0, \nonumber \\
\chi_{3}&=B^{I}{_{0a}} \approx 0, &
\chi_{8}&=2\eta^{abc}\partial_{b}e_{Ic} - 2\partial_{b}\Pi^{ab}{_{I}} + \partial^{a}P_{I} \approx 0, \nonumber \\
\chi_{4}&= \Pi^{0a}{_{I}} \approx 0, &
\chi_{9}&= \Pi^{a}{_{I}} - \eta^{abc}B_{Ibc} \approx 0,  \nonumber \\
\chi_{5}&= \partial^{a}e^{I}{_{a}} \approx 0, &
\chi_{10}&= \Pi^{ab}{_{I}} \approx 0.
\label{eq7}
\end{align}
where $q^{I}$ and $P_{J}$ are auxiliary fields satisfying $\{q^{I}(x),P_{J}(y)\}= \delta^{I}{_{J}}\delta^{3}(x-y)$. In this manner, the matrix whose entries are the Poisson brackets among the constraints  (\ref{eq7}) is given by 
\begin{eqnarray*}
\label{eq}
C_{\alpha \beta}=
\left(
  \begin{array}{cccccccccc}
  0    &\,	  \delta^{I}{_{J}}   &\,  0    &\,  0     &\, 0  &\,  0  &\,  0  &\, 0  &\,  0  &\, 0	 	 \\                                                                        
\delta^{I}{_{J}}  &\,   0   &\,  0   &\,  0  &\,   0   &\,   0  	&\,  0	&\,   0  &\,   0    &\,   0 \\                                                                   
0   &\,  0   &\,  0      &\,  \frac{\delta^{a}{_{b}}}{2}\delta^{I}{_{J}}   &\,   0      &\,  0   &\,  0  &\,   0  &\,   0  &\,   0   \\
    0   &\,  0   &\,  -\frac{\delta^{a}{_{b}}}{2}\delta^{I}{_{J}}     &\,   0  &\,  0   &\,  0   &\,  0 &\,  0 &\,  0 &\,  0 \\
0  &\,  0  &\,    0  &\,  0  &\,  0  &\, -\delta^{I}{_{J}}\nabla^{2} 	&\,  0 	&\,  0  &\, \delta^{I}{_{J}}  \partial^{a}     &\,  0 \\
0   &\,  0      &\,  0     &\, 0      &\,  \delta^{I}{_{J}} \nabla^{2}    &\, 0    &\,  0   &\,  0  &\,0   &\, 0 \\
0     &\,   0      &\,  0     &\, 0      &\,  0       &\,   0     &\, 0        &\,  -\delta^{a}_{b}\delta^{I}{_{J}}   \nabla^{2}  &\,   0    &\, - \frac{ \delta^{gd}_{ab} \delta^{I}{_{J} } \partial^{b} }{2} \\
0     &\,  0      &\, 0    &\,  0      &\,   0      &\,  0       &\,    \nabla^{2}\delta^{a}_{b}\delta^{I}{_{J}}      &\,  0     &\,  0     &\, 0 \\
0     &\,  0     &\,  0   &\, 0     &\, -\delta^{I}{_{J}} \partial^{a}     &\,  0    &\, 0     &\, 0     &\, 0      &\, -\eta^{abc}\eta_{IJ}  \\
%0    &\,  0    &\, 0     &\,  0    &\, 0      &\, 0    &\,\frac{1}{2}\delta^{gd}{_{ab}}\delta^{I}{_{J}}     &\,  0    &\, \eta^{abc}\eta_{IJ}     &\, 0\\
0    &\,  0    &\, 0     &\,  0    &\, 0      &\, 0   &\,  \frac{ \delta^{gd}_{ab} \delta^{I}{_{J}} \partial^{b}}{2}  &\,  0  &\,  \eta^{abc}\eta_{IJ}  &\,  0
 \end{array}
\right) \delta^{3}(x-y),
\end{eqnarray*}\\
and its inverse 
\begin{eqnarray*}
\label{eq}
(C_{\alpha \beta}){^{-1}} =
\left(
  \begin{array}{cccccccccc}
  0    &\,    - \delta^{I}{_{J}}   &\,  0    &\,  0     &\, 0  &\,  0  &\,  0  &\, 0  &\,  0  &\, 0	 	 \\                                                                        
\delta^{I}{_{J}}  &\,   0   &\,  0   &\,  0  &\,   0   &\,   0  	&\,  0	&\,   0  &\,   0    &\,   0 \\                                                                   
0   &\,  0   &\,  0      &\,  -2\delta^{a}{_b}\delta^{I}{_{J}}   &\,   0      &\,  0   &\,  0  &\,   0  &\,   0  &\,   0   \\
    0   &\,  0   &\,  2\delta^{a}{_b}\delta^{I}{_{J}}     &\,   0  &\,  0   &\,  0   &\,  0 &\,  0 &\,  0 &\,  0 \\
0  &\,  0  &\,    0  &\,  0  &\,  0  &\, \frac{\delta^{I}{_{J}} }{\nabla^{2}}  &\,  0 	&\,  0  &\,   0  &\,  0 \\
0   &\,  0      &\,  0     &\, 0      &\,   -\frac{\delta^{I}{_{J}} }{\nabla^{2}}    &\, 0    &\,  0   &\,  0  &\,0   &\, \eta^{IJ} \frac{ \eta_{bac} \partial^{a}}{2\nabla^{2}} \\
0     &\,   0      &\,  0     &\, 0      &\,  0       &\,   0     &\, 0        &\,  \frac{\delta^{a}{_b}}{\nabla^{2}}\delta^{I}{_{J}}    &\,   0    &\, 0 \\
0     &\,  0      &\, 0    &\,  0      &\,   0      &\,  0       &\,    -\frac{\delta^{a}{_b}}{\nabla^{2}}\delta^{I}{_{J}}      &\,  0     &\,  \frac{\eta^{IJ} \eta_{acb} \partial^c}{2 \nabla^2}     &\, 0 \\
0     &\,  0     &\,  0   &\, 0     &\, 0     &\,  0    &\, 0     &\, - \frac{\eta^{IJ} \eta_{acb} \partial^c}{2 \nabla^2}     &\, 0      &\, \frac{\eta_{abc}}{2}\eta^{IJ}  \\
0    &\,  0    &\, 0     &\,  0    &\, 0      &\,  -\eta^{IJ}  \frac{ \eta_{bac} \partial^{a}}{2\nabla^{2}}   &\,0  &\,  0  &\,  -\frac{\eta_{abc}}{2}\eta^{IJ}  &\,  0
 \end{array} 
\right)\\
\times \delta^{3}(x-y).
\end{eqnarray*}\\
It is worth to mention, that  the auxiliar fields $q^{I}$  and $P_{J}$ have been added  because they allow us to find the inverse matrix $C^{-1}_{\alpha \beta}$. In fact,  without these auxiliary fields it is not possible to calculate the inverse matrix,  this is a common problem in reducible theories \cite{17, 18}. However, these auxiliar fields  do not contribute to the dynamics of the system  because the Dirac brackets among the auxiliary fields and  the dynamical variables vanish, this is 
\begin{align*}
\{q^{I}(x),P_{J}(y)\}_{D}&=0, &
\{q^{I}(x),e^{J}{_{0}}(y)\}_{D}&=0, &
\{q^{I}(x),\Pi^{0}{_{J}}(y)\}_{D}&=0,\\
\{q^{I}(x),B^{J}{_{0a}}\}_{D}&=0, &
\{q^{I}(x), \Pi^{0a}{_{J}}(y)\}_{D}&=0, &
\{q^{I}(x),e^{J}{_{a}}(y)\}_{D}&=0, \\
\{q^{I}(x),\Pi^{ab}{_{J}}(y)\}_{D}&=0, &
\{q^{I}(x),\Pi^{a}{_{J}}(y)\}_{D}&=0, &
\{q^{I}(x),B^{J}{_{ab}}(y)\}_{D}&=0,\\
\{P_{I}(x), e^{J}{_{0}}(y)\}_{D}&=0, &
\{P_{i}(x), \Pi^{0}{_{J}}(y)\}_{D}&=0, &
\{P_{I}(x), B^{J}{_{0a}}(y)\}_{D}&=0,\\
\{P_{I}(x), \Pi^{0a}{_{J}}(y)\}_{D}&=0, &
\{P_{I}(x), e^{J}{_{a}}(y)\}_{D}&=0, &
\{P_{I}(x), \Pi^{a}{_{J}}(y)\}_{D}&=0,\\
\{P_{I}(x), \Pi^{ab}{_{J}}(y)\}_{D}&=0.
\end{align*}
After a long computation, we can obtain the following  non-zero Dirac`s brackets  among dynamical variables  given by 
\begin{eqnarray}
\{e^{I}{_{a}}, \Pi^{b}{_{J}}\}_D&=& \delta^{I}{_{J}}\left( \delta^{b}{_{a}} - \frac{\partial _{a}\partial^{b}}{ \nabla ^2} \right) \delta^3(x-y), \nonumber \\
\{e^{I}{_{a}}, B_{Jbc}\}_D&=&\frac{ \delta^{I}{_{J}}}{2} \left( \eta_{abc} - \eta_{dbc}\frac{\partial _{a}\partial^{d}}{ \nabla ^2} \right) \delta^3(x-y). 
\end{eqnarray}
It is important to comment, that all these results  were  not reported in previous works  \cite{13, 14}. On the other hand,  with all these results obtained along this section  we have  at hand all  the necessary tools for comparing at Hamiltonian level  the theory under study and  $BF $ versions of [GR]. In fact, we have comment that the action  (\ref{eq2}) is a copy of a  $BF$ theory, however, the context of these two theories is not the same. For the former we have to Minkowski spacetime as scenario, for the lather there is not scenario at all  because  $BF$ theory  is a  background  independent  theory.  \\
In the following lines we will reproduce the results obtained in this section by means the [FJ] framework.\\
\section{Generalized Faddeev-Jackiw brackets by using the phase space as symplectic variables}
Now, in this section we will reproduce by means of  the [FJ] formulation   the results obtained in the previous section  where the Dirac brackets have been obtained by fixing the gauge. In particular by fixing the gauge we have choosen a particular configuration of the  fields, in this manner,  in [FJ] we should  to work with the phase space as symplectic variables \cite{12}.  In order to perform our analysis, from (\ref{eq8}) we identify the symplectic Lagrangian 
\begin{eqnarray}
\overset{(0)}{\mathcal{L}}&=& \Pi^{a}{_{I}}\dot e^{I}{_{a}} - V^{(0)},
\label{eq14}
\end{eqnarray}
where  $\Pi^{a}{_{I}}= \eta^{abc}B_{Ibc} $ and  $V^{(0)}=-\eta^{abc}B_{I0a}(\partial_{b}e^{I}{_{c}} - \partial_{c}e^{I}{_{b}}) - \partial_{a}\Pi^{a}{_{I}}e^{I}{_{0}}$. Thus, the symplectic coordinates ${\overset{(0)} \varepsilon} ^{i}=(e^{I}{_{a}},\Pi^{a}{_{I}},e^{I}{_{0}},B_{I0a})$ \, and \, $\overset{(0)} a_{i}=(\Pi^{a}{_{I}},0,0,0)$, hence the symplectic matrix is given by 
\begin{eqnarray*}
\label{eq}
\overset{(0)}{f}_{ij}=
\left(
  \begin{array}{ccccc}
  0   &\quad   \delta^{b}{_{a}}\delta^{I}{_{J}}   &\quad   0     &\quad   0                                                                        \\
 -\delta^{b}{_{a}}\delta^{I}{_{J}}   &\quad   0   &\quad   0   &\quad   0                                                                        \\
0   &\quad  0   &\quad   0      &\quad   0                                                                     \\
    0   &\quad   0   &\quad   0    &\quad  0                                                                  \\ 
 \end{array}
\right) \delta^3(x-y), 
\end{eqnarray*}
this matrix has two null vectors given by $ V^{(1)}=(0,0,V^{{e}^{J}{_{0}}},0)$,\quad $V^{(2)}=(0,0,0,V^{{B}_{I0a}})$,
where $V^{{B}_{I0a}}$, \, $V^{{e}^{J}{_{0}}}$ are arbitrary functions. In this manner, by performing the contraction of the null vectors with
\begin{eqnarray}
V^{i}{_{\mu}}\frac{\delta V^{(0)}}{\delta \varepsilon^{\mu}}&=&0,
\label{eq16}
\end{eqnarray}
this implies that  $\overset{(1)}{\Omega}=\partial_{a}\Pi^{a}{_{I}}=0$, \,$\overset{(2)} {\Omega}=\eta^{abc}\partial_{b}e_{Ic}=0$. We are able to observe that these constraints are the secondary constraints found in Dirac's  method. Furthermore, $\overset{(2)}{\Omega}$ is a reducible constraint, because  \, $ \partial_{a} \overset{(2)}{\Omega}{^{a}}=0$,  and we will take  into account  this fact in the following computations. On the other hand, we need calculate the following

\begin{eqnarray*}
\label{eq}
{f}_{ij}=
\left(
 \begin{array}{ccccc}
   \overset{(0)}{f}_{ij}                              \\ 
    \frac{\delta \Omega_{i}}{\delta \varepsilon_{j}}        
\end{array}
\right)
=
\left(
  \begin{array}{ccccc}
   0   & \quad  \delta^{b}{_{a}}\delta^{I}{_{J}}   &\quad  0     & \quad  0                                                                        \\
   -\delta^{b}{_{a}}\delta^{I}{_{J}}   & \quad  0   &\quad  0   &\quad   0                                                                        \\
0   &\quad 0   &\quad   0      &\quad   0                                                                     \\
    0   & \quad  0   &\quad   0    &\quad 0                                                                  \\ 
\eta^{acd}\partial_{c}\delta^{I}{_{J}} &\quad  0 &\quad  0  &\quad  0			\\
0 &\quad  \partial_{a}\delta^{I}{_{J}}  &\quad   0   &\quad  0
  \end{array}
\right) \delta^{3}(x-y).
\end{eqnarray*} \\
We can observe,    this matrix is not squared, however, it has two modes given by 
$(V^{(1)})_{i}^T=(\partial_{b}V^{\lambda}{_{I}},0,V^{{A}_{0}},V_{a}^{B0a},0,-V^{\lambda}{_{I}})$ and
$(V^{(2)})_{i}^{T}= (0,\eta^{bcd}\partial_{c}V_{{dJ}},V^{{A}_{0}},V^{B0a}{_{a}},V_{dJ},0)$. By performing the contraction with the modes, we find that $(V)_{k}^{T}Z_{k}=0$,  is an identity. Therefore, there are not more constraints for the theory under study. In this manner, we will  add all that information by constructing a new symplectic Lagrangian given by 
\begin{eqnarray}
\overset{(1)}{\mathcal{L}}&=& \Pi^{a}{_{I}}\dot e^{I}{_{a}} -2\dot \lambda^{I}{_{a}}(\eta^{abc}\partial_{b}e_{Ic})-\dot \rho^{I}(\partial_{a}\Pi^{a}{_{I}})- \dot \theta_{I}\partial^{a}\lambda^{I}{_{a}}, 
\label{eq22a}
\end{eqnarray}
because of $\overset{(2)} {\Omega}$ is a reducible constraint  in (\ref{eq22a})  was  necessary to add a Lagrange  multiplier $\theta_ I$ of the Lagrange multiplier  $\lambda^I$ \cite{19}. Thus, from  (\ref{eq22a}) we identify the  new set of symplectic variables are given by
$\overset{(1)}{\varepsilon}{^{i}}= \left(e^{I}{_{a}},\Pi^{a}{_{I}},\lambda^{I}{_{a}},\rho^{I},\theta_{I}\right)$ and the 1-forms   $a_{i}=(\Pi^{a}{_{I}},0,-2\eta^{abc}\partial_{b}e_{Ic},-\partial_{a}\Pi^{a}{_{I}},\partial^{a}\lambda^{I}{_{a}})$ in this manner the symplectic matrix takes the form

\begin{eqnarray}
\label{eq}
\overset{(1)}{f}_{ij}=
\left(
  \begin{array}{ccccc}
  0   &\quad   -\delta^{b}{_{a}}\delta^{I}{_{J}}   &\quad  -2\eta^{abc}\eta_{IJ}\partial_{b}     &\quad   0	  & \quad 0 \\                                                                        
 \delta^{a}{_{b}}\delta^{I}{_{J}}   &\quad   0   &\quad  0   &\quad   -\delta^{I}{_{J}}\partial_{a} &\quad   0    \\                                                                   
2\eta_{IJ}\eta^{abc}\partial_{b}   &\quad  0   &\quad   0      &\quad   0   &\quad    -\delta^{I}{_{J}}\partial_{a}    \\
    0   &\quad   \delta^{I}{_{J}}\partial_{a}   &\quad   0    &\quad 0  &\quad  0 \\
0  &\quad  0  &\quad    \delta^{I}{_{J}}\partial^{a}  &\quad  0  &\quad  0 \\
 \end{array}
\right) \delta^{3}(x-y).
\end{eqnarray}\\
We can observe that $\overset{(1)}{f}{_{ij}}$ is singular, however, we have showed that there are not more constraints, the noninvertibility of $\overset{(1)}{f}{_{ij}}$ means that the theory has a gauge simmetry. In this manner, we will take the following fixing gauge\, $\overset{(3)}{\Omega}= \partial^{a}e^{I}{_{a}}=0$ \, and\, $ \overset{(4)}{\Omega}=\frac{1}{2}\eta_{abc}\partial^{b}\Pi^{c}{_{I}}$,  this information needs to be added to the symplectic Lagrangian through one  new Lagrange multipliers, namely, $\eta_I$ and $\alpha^I$. We observe, however, that $\overset{(4)}{\Omega}$  is a reducible constraint (gauge fixing), in this manner in  the symplectic Lagrangian we will add one more  Lagrange multiplier,  $\beta _I$, again, the Lagrange multiplier of the Lagrange multiplier  \cite{19}
\begin{equation}
\begin{split}
\overset{(2)}{\mathcal{L}}&=\Pi^{a}{_{I}}\dot e^{a}{_{I}} - 2\dot \lambda^{I}{_{a}}(\eta^{abc}\partial_{b}e_{Ic}) - \dot \rho^{I}(\partial_{a}\Pi^{a}{_{I}}) - \dot \theta_{I}\partial^{a}\lambda^{I}{_{a}} - (\frac{1}{2}\eta_{abc}\partial^{b}\Pi^{c}{_{I}})\dot \alpha^{aI} \\
&\quad - (\partial^{a}e^{I}{_{a}})\dot \eta_{I} -(\partial^{a}\alpha^{I}{_{a}})\dot{\beta}_{I}. 
\end{split}
\label{eq24}
\end{equation}
Now, from (\ref{eq24}) it is possible to identify the  new set of symplectic variables $\overset{(2)}{\varepsilon}{^{i}}=(e^{a}{_{I}},\Pi^{I}{_{a}},\lambda^{I}{_{a}},\rho^{I},\theta_{I},\alpha^{aI},\eta_{I},\beta_{I}) $ and the  1-form 
$\overset{(2)}a_{i}= (\Pi^{a}{_{I}},0,-2\eta^{abc}\partial_{b}e_{Ic},-\partial_{a}\Pi^{a}{_{I}},-\partial^{a}\lambda^{I}{_{a}}, \frac{1}{2}\eta_{abc}\partial^{b}\Pi^{c}{_{I}},-\partial^{a}e^{I}{_{a}},\partial^{a}\alpha^{I}{_{a}})$. In this manner, the symplectic matrix is given by 
\begin{eqnarray*}
\label{eq}
\overset{(2)}{f}_{ij}=
\left(
  \begin{array}{cccccccc}
  0    &	\quad   -\delta^{b}{_{a}}\delta^{I}{_{J}}   &\quad  -2\eta^{abc}\eta_{IJ}\partial_{b}   &\quad  0     & \quad 0  &\quad  0  &\quad  -\delta^{I}{_{J}} \partial^{a} &\quad 0	 	 \\                                                                        
 \delta^{b}{_{a}}\delta^{I}{_{J}}   &\quad   0   &\quad  0   &\quad   -\delta^{I}{_{J}}\partial_{a}  &\quad   0   &\quad   -\frac{\delta^{I}{_{J}} }{2}\eta_{abc}\partial^{c} &\quad  0	&\quad   0 \\                                                                   
2\eta_{IJ}\eta^{abc}\partial_{b}   &\quad  0   &\quad   0      &\quad   0   &\quad    -\delta^{I}{_{J}}\partial^{a} &\quad  0 	&\quad  0  &\quad   0   \\
    0   &\quad   \delta^{I}{_{J}}\partial_{a}   &\quad   0    &\quad 0  &\quad  0	 &\quad  0	 &\quad  0	 &\quad  0 \\
0  &\quad  0  &\quad    -\delta^{I}{_{J}}\partial^{a}  &\quad  0  &\quad  0 	&\quad  0 	&\quad  0 	&\quad  0 \\
0   &\quad  \frac{ \delta^{I}{_{J}} }{2}\eta_{abc}\partial^{c}  &\quad  0  &\quad  0   &\quad  0 	&\quad  0 	  &\quad  0   &\quad\eta_{IJ}  \partial^{a} \\
\delta^{I}{_{J}}  \partial^{a} &\quad   0   &\quad  0  &\quad  0   &\quad  0 	&\quad  0  &\quad   0     &\quad  0 \\
0  &\quad   0   &\quad  0  &\quad  0   &\quad  0 	&\quad  -\eta_{IJ} \partial^{a}  &\quad   0     &\quad  0 \\
 \end{array}
\right) \\
\times \delta^3(x-y),
\end{eqnarray*}\\
we can observe  that  this matrix is not singular,  thus,  there exists its inverse. The inverse of the symplectic matrix $\overset{(2)}{f}_{ij}$ is given by 
\begin{eqnarray}
\label{eq}
(\overset{(2)}{f}_{ij})^{-1}=
\left(
  \begin{array}{cccccccc}
  0    &\,	  \delta^{I}{_{J}}\left( \delta^{a}{_{b}} - \frac{\partial ^{a}\partial_{b}}{ \nabla ^2} \right)   &\,  0   &\,  0     & \, 0  &\,  0  &\,  \frac{\delta^{I}{_{J}}\partial_{a}}{\nabla^2}  &\,0	 	 \\                                                                        
-\delta^{I}{_{J}}\left( \delta^{a}{_{b}} - \frac{\partial ^{a}\partial_{b}}{ \nabla ^2} \right)  &\,   0   &\,  0   &\,   \frac{\delta^{I}{_{J}}\partial^{a}}{\nabla^2}  &\,   0   &\,  0  &\,  0	&\,   0 \\                                                                   
0   &\,  0   &\,   0      &\,   0   &\,    \frac{\delta^{I}{_{J}}\partial^{a}}{\nabla^2}      &\,  0 	&\,  0  &\,   0   \\
    0   &\,  \frac{-\delta^{I}{_{J}}\partial_{a}}{\nabla^2}   &\,   0    &\,   0  &\,  0	 &\,  0	 &\,  \frac{\delta^{I}{_{J}}}{\nabla^2}	 &\,  0 \\
0  &\,  0  &\,    \frac{-\delta^{I}{_{J}}\partial^{a}}{\nabla^2}  &\,  0  &\,  0 	&\,  0 	&\,  0 	&\,  0 \\
0   &\,   0   &\,  0  &\,  0   &\   0 	&\,  0 	  &\,  0   &\,  \frac{-\delta^{I}{_{J}}\partial_{a}}{\nabla^2} \\
\frac{-\delta^{I}{_{J}}\partial_{a}}{\nabla^2}  &\,   0   &\,  0  &\,  \frac{-\delta^{I}{_{J}}}{\nabla^2}   &\,  0 	&\,  0  &\,   0     &\,  0 \\
0  &\,   0   &\,  0  &\,  0   &\,  0 	&\, \frac{\delta^{I}{_{J}} \partial_{a}}{\nabla^2}  &\,   0     &\,  0 \\
 \end{array}
\right) \nonumber  \\ \times  \delta^{3}(x-y).
\label{eq25a}
\end{eqnarray}\\
Therefore, from (\ref{eq25a}) it is possible to identify the following [FJ]  generalized brackets 
\begin{eqnarray}
\{\xi_{i}^{(2)}(x),\xi_{j}^{(2)}(y)\}_{FD}=[f^{(2)}_{ij}(x,y)]^{-1},
\label{eq28}
\end{eqnarray}
thus
\begin{eqnarray}
\{e^{a}_{I}(x),\Pi^{J}_{b}(y)\}_{FD}=[f^{(2)}_{12}(x,y)]^{-1}=\delta^{I}{_{J}}\left( \delta^{a}{_{b}} - \frac{\partial ^{a}\partial_{b}}{ \nabla ^2} \right) \delta^3(x-y),
\label{eq30}
\end{eqnarray}
that correspond  to the Dirac brackets found in previous section.  It is important to comment that in [FJ] method we did not use auxiliary fields but we used the Lagrange multiplier of the Lagrange multipler \cite{19}, in this sense, the [FJ] framework is more economic than Dirac's method.   \\ 
\section{ Conclusions}
In this paper a detailed  Hamiltonian and a [FJ]  analysis for Einstein's theory in the  $G \rightarrow 0$ limit have    been performed. We have worked  with  the full phase space and we have constructed the Dirac brackets by means two ways,  fixing and without  fixing the gauge.  In this respect,  a complete Hamiltonian study has been performed,  in order to construct the Dirac brackets by fixing the gauge, we have showed that  it is necessary to extend the phase space by means auxiliary variables, however,  this   is a large task   and for this reason we worked with the [FJ] method. In this manner, we have performed a complete [FJ] framework for the theory under study, we  found all the [FJ] constraints and we have constructed the generalized [FJ] brackets;  we have worked with both  the configuration space  and the phase space. In both cases we showed the equivalence between the Dirac and the generalized [FJ] brackets. It is important to comment, that in [FJ]  is not necessary extend the phase space by introducing  canonical auxiliary variables as   Dirac's method requires; in the  [FJ]  scheme  the symplectic matrix is inverted by constraining the Lagrange multipliers. In this sense, the [FJ] framework is more economic  than Dirac's method. Finally, with the results obtained in this paper  we can extend our study to  $BF$  theories \cite{6}  and  pure gravity \cite{8}. In fact, in these papers  a pure Dirac's  analysis  has been performed by working with the complete phase space,  but the Dirac brackets were not calculated; thus,  the learned in this paper  could be useful for that aim. However, all these ideas are in progress and will be reported  in forthcoming works \cite{20}.   \\
 \newline
\newline
\noindent \textbf{Acknowledgements}\\[1ex]
This work was supported by CONACyT under Grant No. CB-2014-01/ 240781. Alberto Escalante wishes to  thank  Eric Gourgoulhon and  the observatoire  de Paris (LUTH)  for the  hospitality.
%%%%%%%%%%%%%%%%%%%%%%%%%%%%%%%%%%%%%%%%%%%%%%%%%%%%%

\end{document}